\documentclass[10pt,letterpaper]{article}
\usepackage[margin=1in,letterpaper]{geometry}
\usepackage{authblk}
\usepackage{graphicx}
\usepackage{amssymb,amsfonts,amsmath}

\usepackage{hyperref}
\usepackage{breakurl} 

\usepackage[numbers,sort&compress]{natbib}

\usepackage{microtype}
\DisableLigatures[f]{encoding = *, family = * }

\usepackage[aboveskip=1pt,labelfont=bf,labelsep=period,singlelinecheck=off]{caption}

\newcommand{\beginsupplement}{%
        \setcounter{table}{0}
        \renewcommand{\thetable}{S\arabic{table}}%
        \setcounter{figure}{0}
        \renewcommand{\thefigure}{S\arabic{figure}}%
     }
     
\title{The Distribution of the Asymptotic Number of Citations to Sets of Publications by a Researcher or From an Academic Department Are Consistent With a Discrete Lognormal Model}

\author[1]{Jo\~ao A. G. Moreira}
\author[1]{Xiao Han T. Zeng}
\author[1,2,3,4]{Lu\'is A. Nunes Amaral}

\affil[1]{Department of Chemical and Biological Engineering, Northwestern University, Evanston, IL, USA}
\affil[2]{Department of Physics and Astronomy, Northwestern University, Evanston, IL, USA}
\affil[3]{Northwestern Institute on Complex Systems, Northwestern University, Evanston, IL, USA}
\affil[4]{Howard Hughes Medical Institute, Northwestern University, Evanston, IL, USA}

\date{}

\begin{document}

\maketitle

\section*{Abstract}
How to quantify the impact of a researcher's or an institution's body of work is a matter of increasing importance to scientists, funding agencies, and hiring committees. The use of bibliometric indicators, such as the \emph{h}-index or the Journal Impact Factor, have become widespread despite their known limitations. We argue that most existing bibliometric indicators are inconsistent, biased, and, worst of all, susceptible to manipulation. Here, we pursue a principled approach to the development of an indicator to quantify the scientific impact of both individual researchers and research institutions grounded on the functional form of the distribution of the asymptotic number of citations. We validate our approach using the publication records of 1,283 researchers from seven scientific and engineering disciplines and the chemistry departments at the 106 U.S. research institutions classified as ``very high research activity''. Our approach has three distinct advantages. First, it accurately captures the overall scientific impact of researchers at all career stages, as measured by asymptotic citation counts. Second, unlike other measures, our indicator is resistant to manipulation and rewards publication quality over quantity. Third, our approach captures the time-evolution of the scientific impact of research institutions.

\section*{Introduction}
The explosive growth in the number of scientific journals and publications has outstripped researchers' ability to evaluate them \cite{Nicholas2014}. To choose what to browse, read, or cite from a huge and growing collection of scientific literature is a challenging task for researchers in nearly all areas of Science and Technology. In order to search for worthwhile publications, researchers are thus relying more and more on heuristic proxies -- such as author and journal reputations -- that signal publication quality.

The introduction of the \textit{Science Citation Index} (SCI) in 1963 \cite{Garfield1963} and the establishment of bibliographic databases spurred the development of bibliometric measures for quantifying the impact of individual researchers, journals, and institutions. Various bibliometric indicators have been proposed as measures of impact, including such notorious examples as the Journal Impact Factor and the \emph{h}-index \cite{Garfield2006, Hirsch2005}. However, several studies revealed that these measures can be inconsistent, biased, and, worst of all, susceptible to manipulation \cite{MacRoberts1989, Narin1996, Cole2000, Glanzel2002, Borgman2002, Vinkler2004, Bornmann2007, Bornmann2008, Alonso2009, Castellano2009, Wilhite2012}. For example, the limitations of the popular \emph{h}-index include its dependence on discipline and on career length \cite{Egghe2007}.

In recent years, researchers have proposed a veritable alphabet soup of ``new'' metrics -- the \emph{g}-index \cite{Egghe2006b}, the \emph{R}-index \cite{Jin2007a}, the \emph{ch}-index \cite{Franceschini2010}, among others -- most of which are \textit{ad-hoc} heuristics, lacking insight about why or how scientific publications accumulate citations.

The onslaught of dubious indicators  based on citation counts has spurred a backlash and the introduction of so-called ``altmetric'' indicators of scientific performance. These new indicators completely disregard citations, considering instead such quantities as number of article downloads or article views, and number of ``shares'' on diverse social platforms \cite{Bonetta2009, Fausto2012, Kwok2013}. Unfortunately, new research is showing that altmetrics are likely to reflect popularity rather than impact, that they have incomplete coverage of the scientific disciplines \cite{Haustein2011, Priem2012}, and that they are \textit{extremely susceptible to manipulation}. For example, inflating the findings of a publication in the abstract can lead to misleading press reports \cite{Yavchitz2012}, and journals' electronic interfaces can be designed to inflate article views and/or downloads \cite{Davis2006}.
    
Citations are the currency of scientific research. In theory, they are used by researchers to recognize prior work that was crucial to the study being reported. However, citations are also used to make the research message more persuasive, to refute previous work, or to align with a given field \cite{Brooks1986}. To complicate matters further, the various scientific disciplines differ in their citation practices \cite{Rosvall2008}. Yet, despite their limitations, citations from articles published in reputable journals remain the most significant quantity with which to build indicators of scientific impact \cite{Bornmann2008}.

It behooves us to develop a measure that is based on a thorough understanding of the citation accumulation process and also grounded on a rigorous statistical validation. Some researchers have taken some steps in this direction. Examples include the ranking of researchers using PageRank \cite{Radicchi2009} or the beta distribution \cite{Petersen2011}, and the re-scaling of citation distributions from different disciplines under a universal curve using the lognormal distribution \cite{Radicchi2008}.

One crucial aspect of the process of citation accumulation is that it takes a long time to reach a steady state \cite{Stringer2008}. This reality is often ignored in many analyses and thus confounds the interpretation of most measured values. Indeed, the lag  between time of publication and perception of impact is becoming increasingly relevant. For example, faced with increasingly large pools of applicants, hiring committees need to be able to find the most qualified researchers for the position in an efficient and timely manner \cite{Lehmann2006, Abbott2010}. To our knowledge, only a few attempts have been made in developing indicators that can predict future impact using citation measures \cite{Acuna2012, Mazloumian2012} and those have had limited success \cite{Penner2013}.

Here, we depart from previous efforts by developing a principled approach to the quantification of scientific impact. Specifically, we demonstrate that the distribution of the asymptotic number of accumulated citations to publications by a researcher or from a research institution is consistent with a discrete lognormal model \cite{Stringer2008, Stringer2010}. We validate our approach with two datasets acquired from Thomson Reuters' Web of Science (WoS):
\begin{itemize}
\item Manually disambiguated citation data pertaining to researchers at the top United States (U.S.) research institutions across seven disciplines \cite{Duch2012}: chemical engineering, chemistry, ecology, industrial engineering, material science, molecular biology, and psychology;
\item Citation data from the chemistry departments of 106 U.S. institutions classified as ``very high research activity''.
\end{itemize}
Significantly, our findings enable us to develop a measure of scientific impact with desirable properties.

\section*{The Data}
We perform our first set of analyses on the dataset described by Duch et al.\ \cite{Duch2012}. This dataset contains the disambiguated publication records of 4,204 faculty members at some of the top U.S. research universities in seven scientific disciplines: chemical engineering, chemistry, ecology, industrial engineering, material science, molecular biology, and psychology (see \cite{Duch2012} for details about data acquisition and validation). We consider here only 230,964 publications that were in press by the end of 2000. We do this so that every publication considered has had a time span of at least 10 years for accruing citations \cite{Stringer2010} (the researcher's publication dataset was gathered in 2010).

We perform our second set of analyses on the publication records of the chemistry departments at the top U.S. research institutions according to \cite{wiki}. Using the publications' address fields, we identified 382,935 total publications from 106 chemistry departments that were in press by the end of 2009 (the department's publication dataset was gathered in 2014).

In our analyses we distinguish between ``primary'' publications, which report original research findings, and ``secondary'' publications, which analyze, promote or compile research published elsewhere. We identify as primary publications those classified by WoS as ``Article'', ``Letter'', or ``Note'' and identify all other publications types as secondary publications.

Moreover, to ensure that we have enough statistical power to determine the significance of the model fits, we restrict our analysis to researchers with at least 50 primary research publications. These restrictions reduce the size of the researchers dataset to 1,283 researchers and 148,878 publications. All 106 departments in our dataset have a total of more than 50 primary research publications.

\section*{The Distribution of the asymptotic Number of Citations}
Prior research suggests that a lognormal distribution can be used to approximate the steady-state citation profile of a researcher's aggregated publications \cite{Redner2005, Radicchi2008}. Stringer et al.\ demonstrated that the distribution of the number $n(t)$ of citations to publications published in a given journal in a given year converges to a stationary functional form after about ten years \cite{Stringer2008}. This result was interpreted as an indication that the publications published in a single journal have a characteristic citation propensity \cite{Burrell2003a} which is captured by the distribution of the ``ultimate'' number of citations. Here, we investigate the asymptotic number of citations $n_a$ to the publications of an individual researcher as well as the set of all researchers in a department at a research institution.

We hypothesize that $n_a$ is a function of a latent variable $\psi$ representing a publication's ``citability'' \cite{Burrell2001}. The citability $\psi$ results from the interplay of several, possibly independent, variables such as timeliness of the work, originality of approach, strength of conclusion, reputation of authors and journals, and potential for generalization to other disciplines, just to name a few \cite{Shockley1957, Letchford2015}. In the simplest case, citability will be additive in all these variables, in which case the applicability of the central limit theorem implies that $\psi$ will be a Gaussian variable, $\psi \in N(\mu_a, \sigma_a)$, where $\mu_a$ and $\sigma_a$ are respectively the mean and standard deviation of the citability of the publications by researcher $a$. Therefore, the impact of a researcher's body of work is described by a distribution characterized by just two parameters, $\mu$ and $\sigma$. Similarly, because in the U.S. departments hire faculty based on their estimated quality, the researchers associated with a department will presumably be similar in stature or potential.

Unlike citations, which are observable and quantifiable, the variables contributing to $\psi$ are neither easily observable nor easy to quantify. Moreover, mapping $\psi$ into citations is not a trivial matter. Citation counts span many orders of magnitude, with the most highly cited publications having tens of thousands of citations \cite{VanNoorden2014}. Large-scale experiments on cultural markets indicate that social interactions often create a ``rich get richer'' dynamics, far distancing the quality of an underlying item from its impact \cite{Salganik2006}. Citation dynamics are no different. For example, Duch et al.\ recently showed that the \emph{h}-index has a power-law dependence on the number of publications $N_p$ of a researcher \cite{Duch2012}. Here, we reduce the potential distortion of citation-accruing dynamics by focusing on the logarithm of $n_a$. In effect, we take $n_a$ to be the result of a multiplicative process of the same variables determining $\psi$. Thus, we can calculate the probability $p_{dln}(n_a)$ that a researcher or department will have a primary research publication with $n_a$ citations, as an integral over $\psi$:
\begin{equation}\label{eq:lognorm-eq}
p_{dln}(n_a|\mu,\sigma) = \int\limits_{\log_{10}(n_a)}^{\log_{10}(n_a+1)}\frac{\mathrm{d}\psi}{\sqrt{2\pi\sigma^2}}\exp{\left(-\frac{(\psi-\mu)^2}{2\sigma^2}\right)}\;\;.
\end{equation}

Most researchers also communicate their ideas to their peers via secondary publications such as conference proceedings which, in many disciplines, are mainly intended to promote related work published elsewhere. Some secondary  publications will have significant timeliness, in particular review papers and editorial materials, and therefore will likely be cited too. Most of them, however, will not be cited at all. If accounting for secondary publications, Eq. \eqref{eq:lognorm-eq} has to be generalized as:
\begin{equation}\label{eq:lognorm-full}
P(n_a|\mu,\sigma,f_s, \boldsymbol{\theta}) = (1-f_s) p_{dln}(n_a|\mu,\sigma) + f_s \, p_s(n_a|\boldsymbol{\theta})\;\;,
\end{equation}
where $f_s$ is the fraction of secondary publications in a body of work and $p_s(n_a| \boldsymbol{\theta})$ represents the probability distribution, characterized by parameters $\boldsymbol{\theta}$ and not necessarily lognormal, of $n_a$ for secondary research publications. We found that in practice Eq.~\eqref{eq:lognorm-full} can be well approximated by:
\begin{equation*}
P(n_a|\mu',\sigma',f_s) = f_s \, \delta_{0,n_a} + (1-f_s) p_{dln}(n_a|\mu',\sigma')\;\;,
\end{equation*}
where $\delta$ is the Kronecker delta. Surprisingly, we found that $\mu' \approx \mu$ and $\sigma' \approx \sigma$, suggesting that secondary publications have citation characteristics that are significantly different from those of primary publications.

\section*{Results}
\begin{figure}[t]
\centerline{\includegraphics[width=0.8\textwidth]{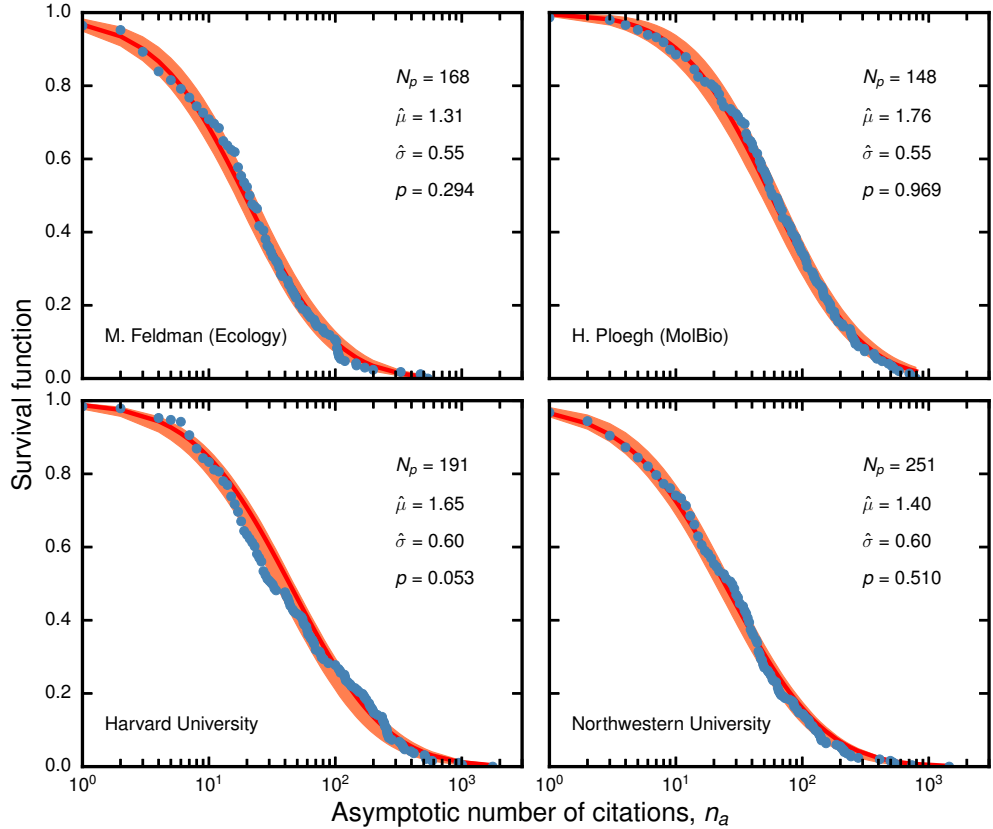}}
\caption{{\bf Distribution of the asymptotic number of citations to publications for researchers and chemistry departments in our database.} We fit Eq.~\eqref{eq:lognorm-eq} to all citations accrued by 2010 to publications published by 2000 for two researchers (\textbf{top row}), and to all citations accrued by 2013 to publications published in 2000 for two chemistry departments (\textbf{bottom row}). The red line shows the maximum likelihood fit of Eq.~\eqref{eq:lognorm-eq} to the data (blue circles). The light red region represents the 95\% confidence interval estimated using bootstrap (1000 generated samples per empirical data point). We also show the number of publications $N_p$ in each set and the parameter values of the individual fits.}
\label{fig1}
\end{figure}

Figure \ref{fig1} shows the cumulative distribution of citations to primary research publications of two researchers in our database and two chemistry departments. Using a $\chi^2$ goodness-of-fit test with re-sampling \cite{Manly2006}, we find that we can reject the discrete lognormal model, Eq.~\eqref{eq:lognorm-eq}, for only 2.88\% of researchers and 1.13\% or departments in our database. The results of our statistical analysis demonstrate that a discrete lognormal distribution with parameters $\mu$ and $\sigma$ provides an accurate description of the distribution of the asymptotic number of citations for a researcher's body of work and for the publications from an academic department.

\begin{figure}[t]
\centerline{\includegraphics[scale=0.3]{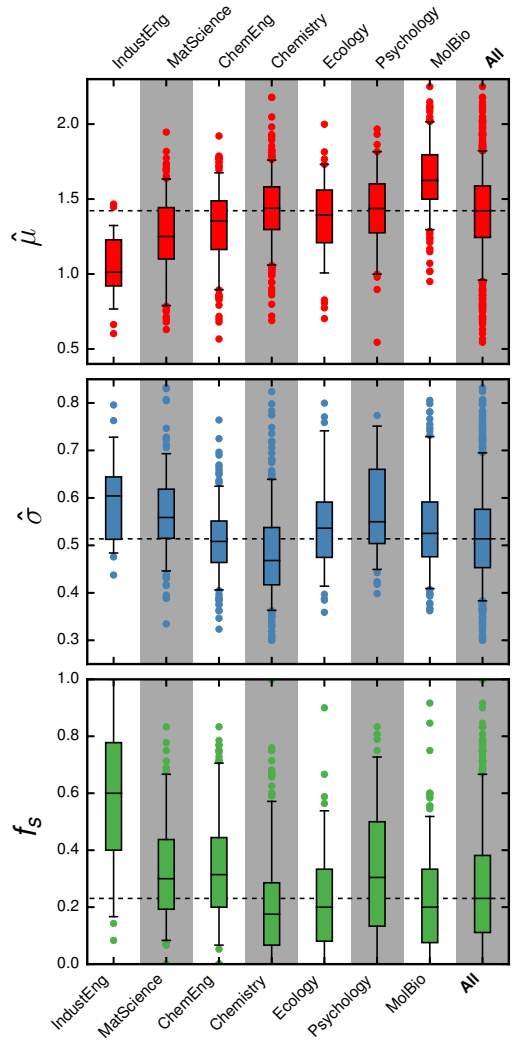}}
\caption{{\bf Parameter statistics of all 1,283 researchers in the database grouped by discipline.} We show the maximum likelihood fitted model parameters (\textbf{top} and \textbf{center}) and the fraction of secondary publications (\textbf{bottom}). The black horizontal dashed line indicates the median of all researchers. For clarity, we do not show the values of $\hat{\sigma}$ for 9 researchers that are outliers.}
\label{fig2}
\end{figure}

\begin{figure}[t]
\centerline{\includegraphics[width=0.6\textwidth]{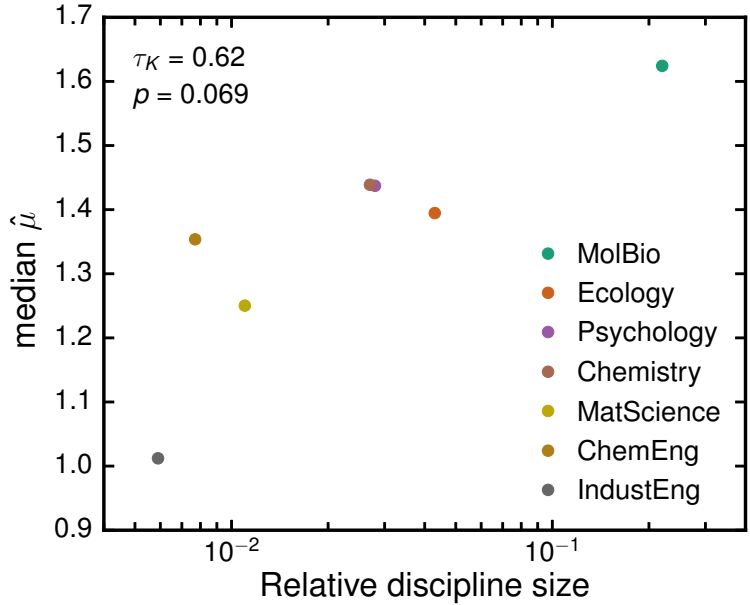}}
\caption{{\bf Correlation between median $\hat{\mu}$ for a discipline and the discipline's relative size.} We use Rosvall et al.\ \cite{Rosvall2008} reported values of the relative number of citations to publications in journals of several disciplines as a proxy for relative field size and compare them with the median value of $\hat{\mu}$ in each discipline. A Kendall rank-correlation test yields a $\tau_K =$ 0.62 with $p =$ 0.069. This correlation suggests that $\hat{\mu}$ depends on the typical number of citations of a discipline.}
\label{fig3}
\end{figure}

Figure \ref{fig2} displays the sample characteristics of the fitted parameters. The median value of $\hat{\mu}$ obtained for the different disciplines lies between 1.0 and 1.6. Using data reported in \cite{Rosvall2008} we find a significant correlation ($\tau_{Kendall}$ = 0.62, \textit{p} = 0.069) between the median value of $\hat{\mu}$ for a discipline and the total number of citation to journals in that discipline (Fig.~\ref{fig3}). This correlation suggests that $\hat{\mu}$ depends on the typical number of citations to publications within a discipline. This dependence on discipline size can in principle be corrected by a normalization factor \cite{Radicchi2008, Castellano2009, Petersen2010}. 

We also plot the fraction of secondary publications, $f_s$, for all the researchers. We find that nearly a fourth of the publications of half of all researchers are secondary, but intra-discipline variation is high. Inter-discipline variability is also high: 17\% of the publications of a typical researcher in chemistry are secondary, whereas 60\% of the publications of a typical researcher in industrial engineering are secondary.

\subsection*{Reliability of Estimation}

We next investigate the dependence of the parameter estimates on number of publications, $N_p$, both at the individual level -- testing the effect of sample size -- and at the discipline level -- testing overall dependence on $N_p$. To test for sample size dependence, we fit the model to subsets of a researcher's publication list. We find that estimates of $\sigma$ are more sensitive to sample size than estimates of $\mu$ (Figs.~\ref{S1_Fig} and \ref{S2_Fig}). However, this dependence becomes rapidly negligible as the sample size approaches the minimum number of publications we required in creating our sample ($N_p \ge$ 50).

Next, we test whether, at the discipline level, there is any dependence of $\hat{\mu}$ on $N_p$. We find no statistically significant correlation, except for a very weak dependence ($\textrm{R}^2 \sim$ 0.035, \textit{p} = 0.0052) of $\hat{\sigma}$ on $N_p$ for chemical engineering (Table \ref{S1_Table}). This is in stark contrast with the \emph{h}-index which exhibits a marked dependence on number of publications \cite{Egghe2007}.

Then, we test for variation of the estimated parameter values along a researcher's career. To this end, we order each researcher's publication records chronologically and divide them into three sets with equal number of publications and fitted the model to each set of publications. Each set represents the citability of the publications authored at a particular career stage of a researcher. Time trends in the estimated values of $\mu$ would indicate that the citability of a researcher's work changes over time. We find such a change for 25\% of all researchers. For over 64\% of those researchers whose citability changes of over time we find that $\hat{\mu}$ increases (Table \ref{table1}).

\begin{table}[t]
\begin{center}
\begin{tabular}{l|r|r}
\hline
{\bf Discipline} & {\bf Upward trend in $\hat{\mu}$} & {\bf Downward trend in $\hat{\mu}$} \\ \hline
ChemEng & 12\% & 19\% \\
Chemistry & 26\% & 6\% \\
Ecology & 8\% & 8\% \\
IndustEng & 0\% & 33\% \\
MatScience & 10\% & 11\% \\
MolBio & 5\% & 8\% \\
Psychology & 0\% & 0\% \\ \hline
\textbf{All} & \textbf{16\%} & \textbf{9\%} \\ \hline
\end{tabular}
\end{center}
\caption{{\bf Trends of $\hat{\mu}$ on career stage for the seven disciplines considered.} We divide each researcher's chronologically-ordered publication records into three sets with equal number of publications (start, middle, and end) and fit the model to each set of publications to obtain $\hat{\mu}^s$, $\hat{\mu}^m$, and $\hat{\mu}^e$. We then used ordinary-least-squares to perform a linear regression on the time dependence of $(\hat{\mu}^s, \hat{\mu}^m, \hat{\mu}^e)$. We then calculate the fraction of researchers whose $\mu$ exhibits a statistically significant dependence on career length, by performing a two-tailed significance test on the slope of the regression. We use a randomization test (1,000 samples), combined with a multiple hypothesis correction \cite{Benjamini1995} (\emph{false discovery rate} of 0.05) to calculate a \textit{p}-value: for each researcher, we randomly re-order his or her publications, divide them into three sets with equal number of publications and fit the model to each set of publications, and calculate the new slope; we obtain a \textit{p}-value by comparing the original slope of the fit with the distribution of the randomized slopes.}
\label{table1}
\end{table}

\begin{figure}[t]
\centerline{\includegraphics[width=\textwidth]{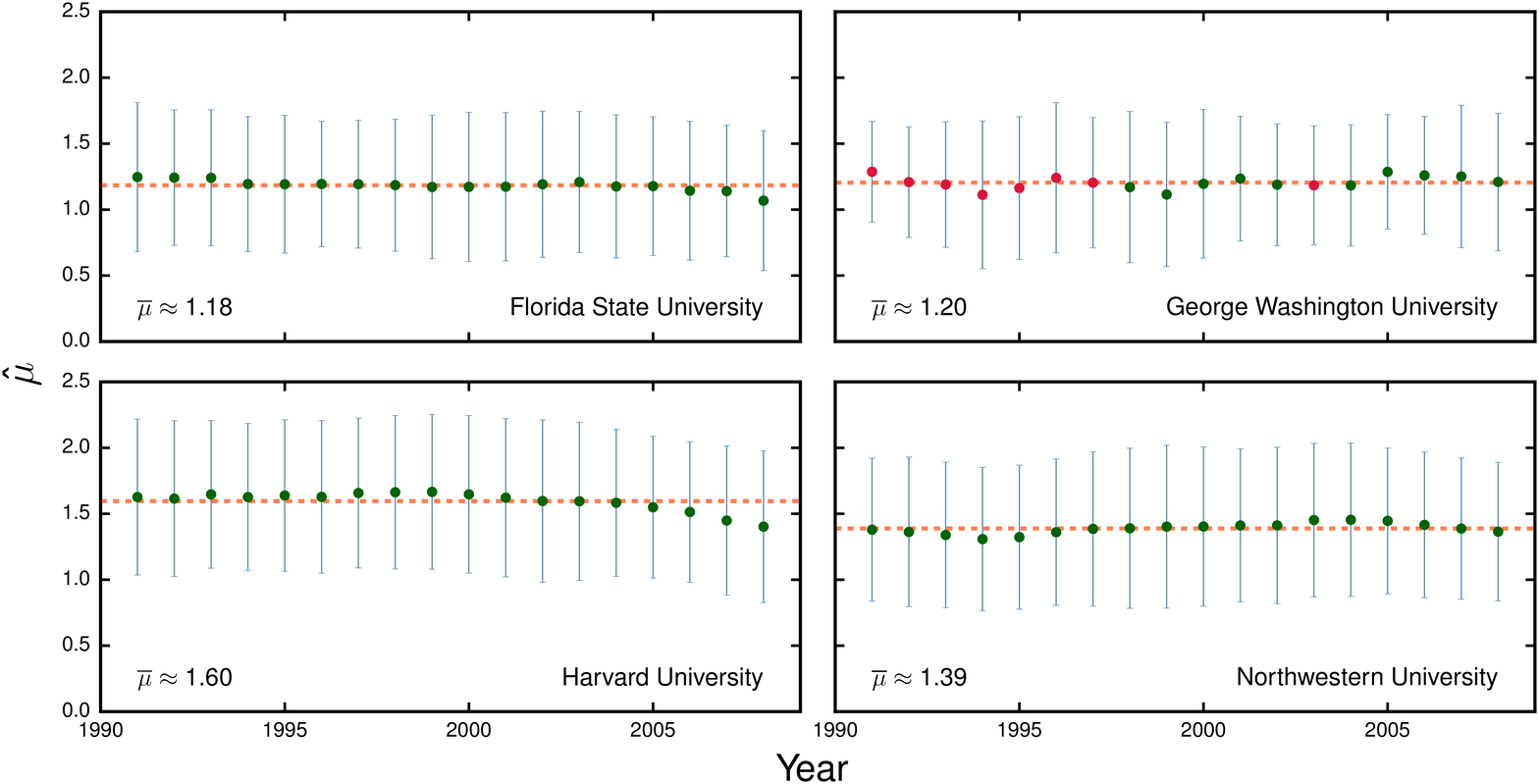}}
\caption{{\bf Time-evolution of departments $\hat{\mu}$.} Each circle and bar represent, respectively, the $\hat{\mu}$ and $\hat{\sigma}$ for a given year of publications. We estimate the parameters in Eq.~\eqref{eq:lognorm-eq} for sets of departmental publications using a ``sliding window'' of 3 years. Fits for which we cannot reject the hypothesis that the data is consistent with a discrete lognormal distribution are colored green. We also show each department's average value of $\hat{\mu}$ over the period considered (orange dashed lines).}
\label{fig4}
\end{figure}

In general, a department has many more publications than any single researcher. Thus, we are able to apply the model from Eq.~\eqref{eq:lognorm-eq} to each year's worth of departmental publications. This fine temporal resolution enables us to investigate whether there is any time-dependence in the citability of the publications from a department. Figure \ref{fig4} shows the time-evolution of $\hat{\mu}$ for the chemistry departments at four typical research institutions. We see that both $\hat{\mu}$ (circles) and $\hat{\sigma}$ (vertical bars) remain remarkably stable over the period considered.

\subsection*{Development of an Indicator}

\begin{figure}[t]
\centerline{\includegraphics[width=0.8\textwidth]{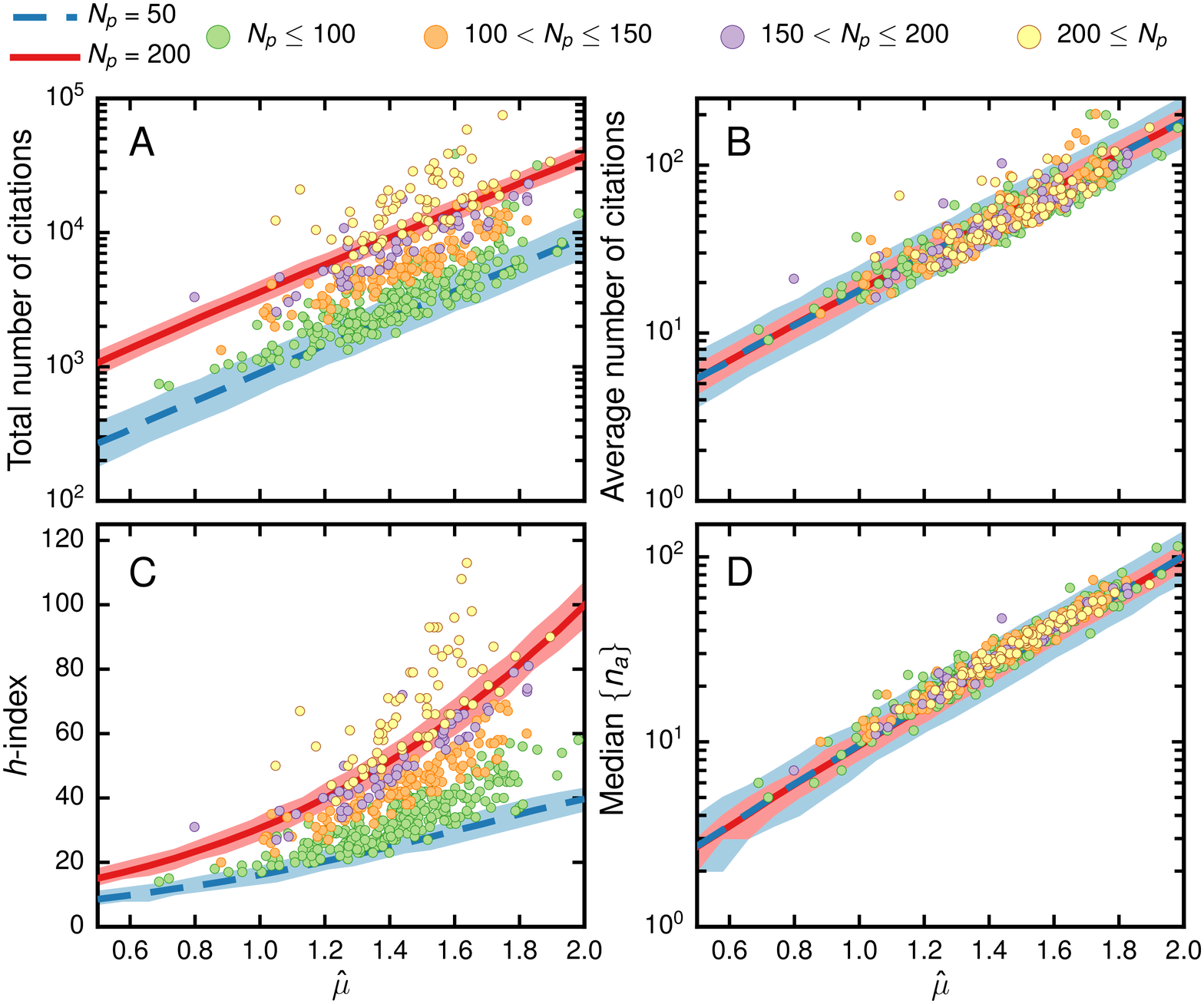}}
\caption{{\bf Dependence of popular impact metrics on the values of $\hat{\mu}$ and number of publications $N_p$ for researchers in chemistry.} We generate 1000 synthetic datasets for each of 20 values of $\hat{\mu}$ from 0.5 to 2.0, inclusive, and for $N_p$ = 50 (blue) and $N_p$ = 200 (red). We use the average $\hat{\sigma}$ of all researchers in chemistry. For each pair of values of $\hat{\mu}$ and $N_p$ we calculated the average value and 95\% confidence interval. The colored circles indicate the observed values of the corresponding metrics for chemistry, which have been grouped according to their number of publications $N_p$. Values for 22 researchers fall outside of the figures' limits: 3 in A, 7 in B, 4 in C, 3 in D. (A) The total number of citations depends dramatically on $N_p$, which in turn depends strongly on career length, and can be influenced by just a few highly cited publications. (B) The average number of citations is less susceptible to changes in $N_p$ but can still be influenced by a small number of highly cited publications. (C) The \emph{h}-index, like the total number of publications, is strongly dependent on $N_p$. (D) The median number of citations to publications, like the average, is not very dependent on $N_p$, and can capture most of the observed behavior.}
\label{fig5}
\end{figure}

In the following, we compare the effectiveness of $\mu$ as an impact indicator with that of other indicators. First, we test the extent to which the value of $\mu_i$ for a given researcher is correlated with the values of other indicators for the same researcher. In order to provide an understanding of how the number of publications $N_p$ influences the values of other metrics, we generate thousands of synthetic samples of $n_a$ for different values of $N_p$ and $\mu_i$, and a fixed value of $\sigma$ for each discipline. We find that $\mu$ is tightly correlated with several other measures, especially with the median number of citations (Fig.~\ref{fig5}). Indeed $\hat{\mu}$ can be estimated from the median number of citations:
\begin{equation} \label{eq:mu-med}
\hat{\mu} \cong \textrm{log}_{10} [\textrm{median}(n_a)]\;\;,
\end{equation}
This close relation between mean and logarithm of the median further supports our hypothesis of a lognormal distribution for the asymptotic number of citations to primary publications by a researcher.

An important factor to consider when designing a bibliometric indicator is its susceptibility to manipulation. Both the number of publications and total or average number of citations are easily manipulated, especially with the ongoing proliferation of journals of dubious reputation \cite{Bohannon2013, Butler2013}. Indeed, the \emph{h}-index was introduced as a metric that resists manipulation. However, it is a straightforward exercise to show that one could achieve $h \propto \sqrt{N_p}$ exclusively through self-citations. Indeed, because the \emph{h}-index does not account for the effect of self-citations, it is rather susceptible to manipulation, especially by researchers with low values of \emph{h} \cite{Schreiber2007,Engqvist2008}.

\begin{figure}[t]
\centerline{\includegraphics[width=0.9\textwidth]{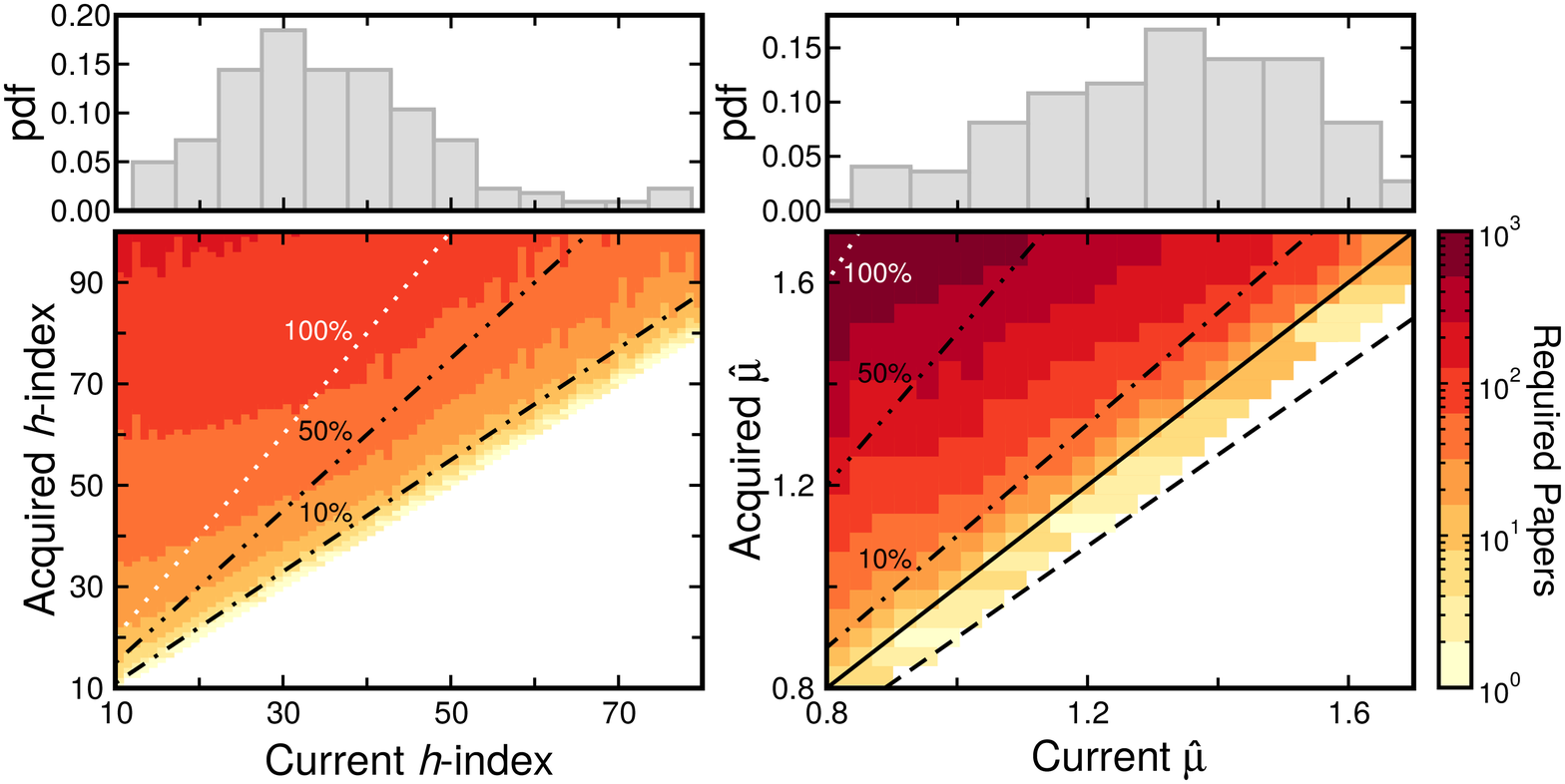}}
\caption{{\bf Comparison of the susceptibility of \emph{h}-index (left) and $\mu$ (right) to manipulation}. \textbf{Bottom panel}: For each researcher in the database, we add publications with self-citations until we reach the desired value of index (see main text for details). The dashed black, dotted-dashed black and dotted white lines indicate the number of publications required to increase the index value by 10\%, 50\% and 100\%, respectively. The solid diagonal black line indicates when the current value of $\hat{\mu}$ is equal to the manipulated $\hat{\mu}$. The dark blue vertical line represents the average value of the indicator amongst all researchers in our database. \textbf{Top panel}: Distributions of current \emph{h}-index (left) and $\hat{\mu}$ (right) for all researchers in the database.}
\label{fig6}
\end{figure}

In order to determine the true susceptibility of the \emph{h}-index to manipulation, we devise a method to raise a researcher's \emph{h}-index using the least possible number of self-citations (see Materials and Methods for details). Our results suggest that increasing the \emph{h}-index by a small amount is no hard feat for researchers with the ability to quickly produce new articles (Fig.~\ref{fig6}A). 

Our proposed indicator, $\mu$, is far more difficult to manipulate. Because it has a more complex dependence on the number of citations than the \emph{h}-index, to increase $\mu$ in an efficient manner we use a process whereby we attempt to increase the median number of citations of a researcher's work (see Materials and Methods for details). Specifically, we manipulated $\mu$ for all the researchers by increasing their median number of citations. Remarkably, to increase $\mu$ by a certain factor one needs at least 10 times more self-citations than one would need in order to increase the \emph{h}-index by the same factor (Fig.~\ref{fig6}B).

While a difference of 2 to 3 orders of magnitude in number of required self-citations may seem surprising for a measure so correlated with citation numbers (Fig.~\ref{fig5}), the fact that $\hat{\mu}$ is actually dependent on the citations to half of all primary publications by a researcher (Eq.~\eqref{eq:mu-med}) makes $\hat{\mu}$ less susceptible than the \emph{h}-index to manipulation of citation counts from a small number of publications. This view is also supported by the fact that increasing citations may actually decrease $\hat{\mu}$, as we may be adding them to a publication that would not be expected to receive that number of citations given the lognormal model. As a result, manipulation of scientific performance would be very difficult if using a $\mu$-based index.

\subsection*{Comparison of Parameter Statistics}

\begin{figure}[t]
\centerline{\includegraphics[width=0.6\textwidth]{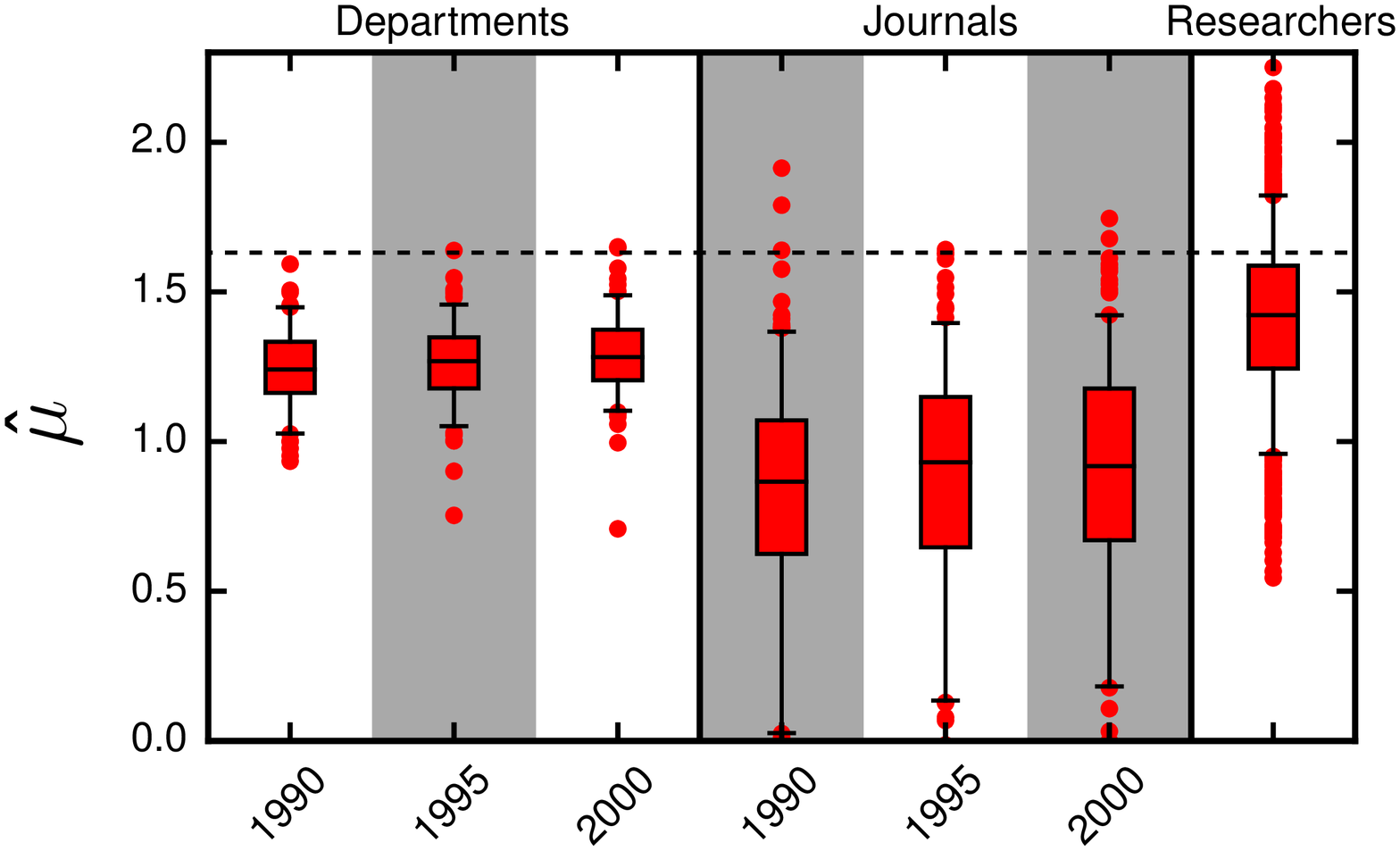}}
\caption{{\bf Comparison of $\hat{\mu}$ across departments, journals, and researchers.} We show the maximum likelihood fitted $\hat{\mu}$ for chemistry departments and chemistry journals in select years, and for all chemistry researchers in our database. The black horizontal dashed lines mark the value of the corresponding parameter for the \textit{Journal of the American Chemical Society} in 1995. For clarity, we do not show $\hat{\mu}$ for 23 journals that are outliers.}
\label{fig7}
\end{figure}

\begin{figure}[t]
\centerline{\includegraphics[width=0.9\textwidth]{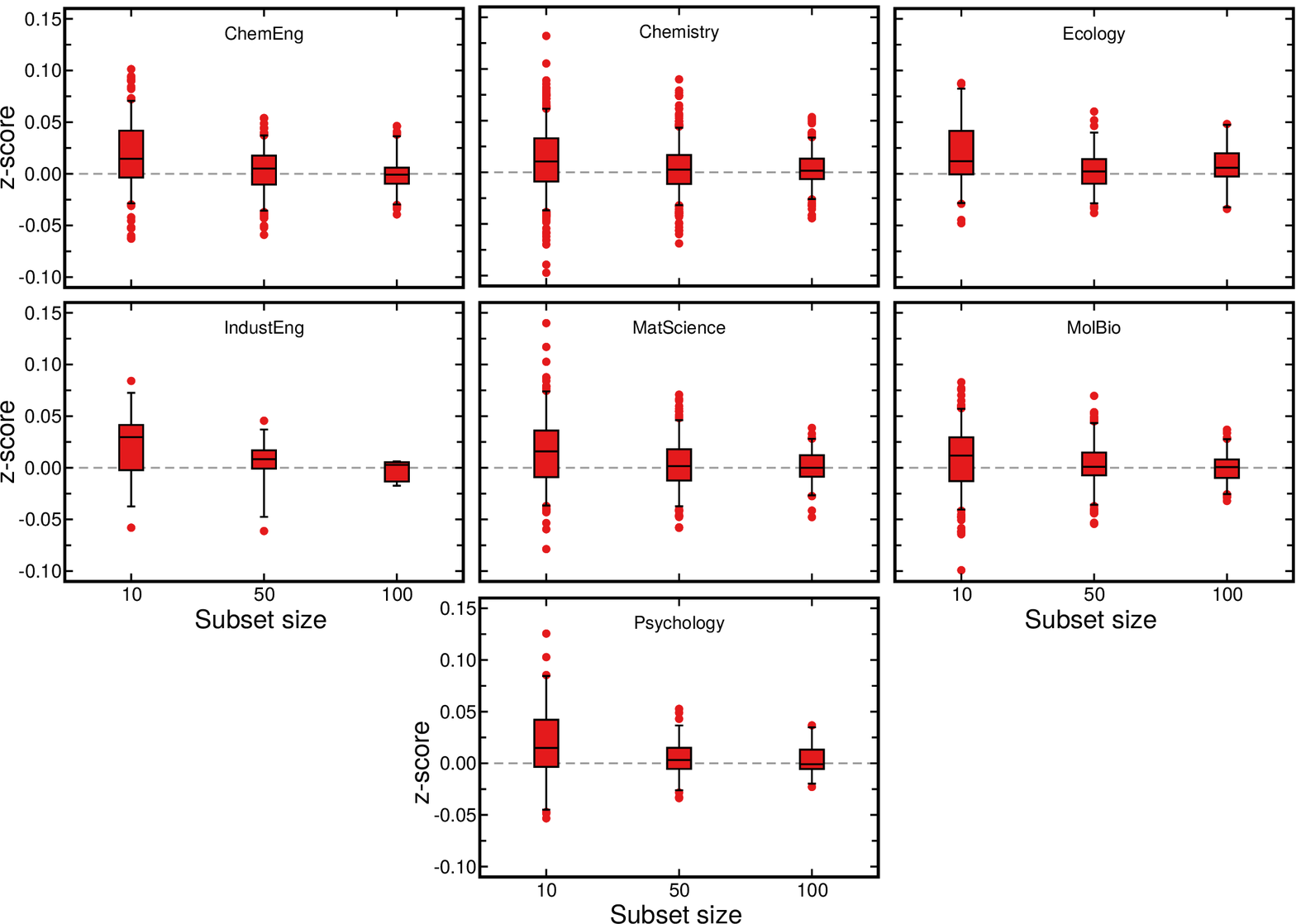}}
\caption{{\bf Dependence of $\hat{\mu}$ on number of publications at the individual level.} We fit the model to 1,000 randomized subsets of each researcher's publication list and compare the $\hat{\mu}$ obtained from fitting each subset of 10, 50, and 100 publications with the $\hat{\mu}$ associated with the complete publication list. Then, for each researcher and subset size, we calculate a z-score using the mean and standard deviation of the ``sub-$\hat{\mu}$''. For $N_p \geq$ 50, the dependence on sample size is negligible for most researchers. Researchers with $N_p <$ 100 are omitted from the calculation on the subset of size 100.}
\label{S1_Fig}
\end{figure}

\begin{figure}[t]
\centerline{\includegraphics[width=0.9\textwidth]{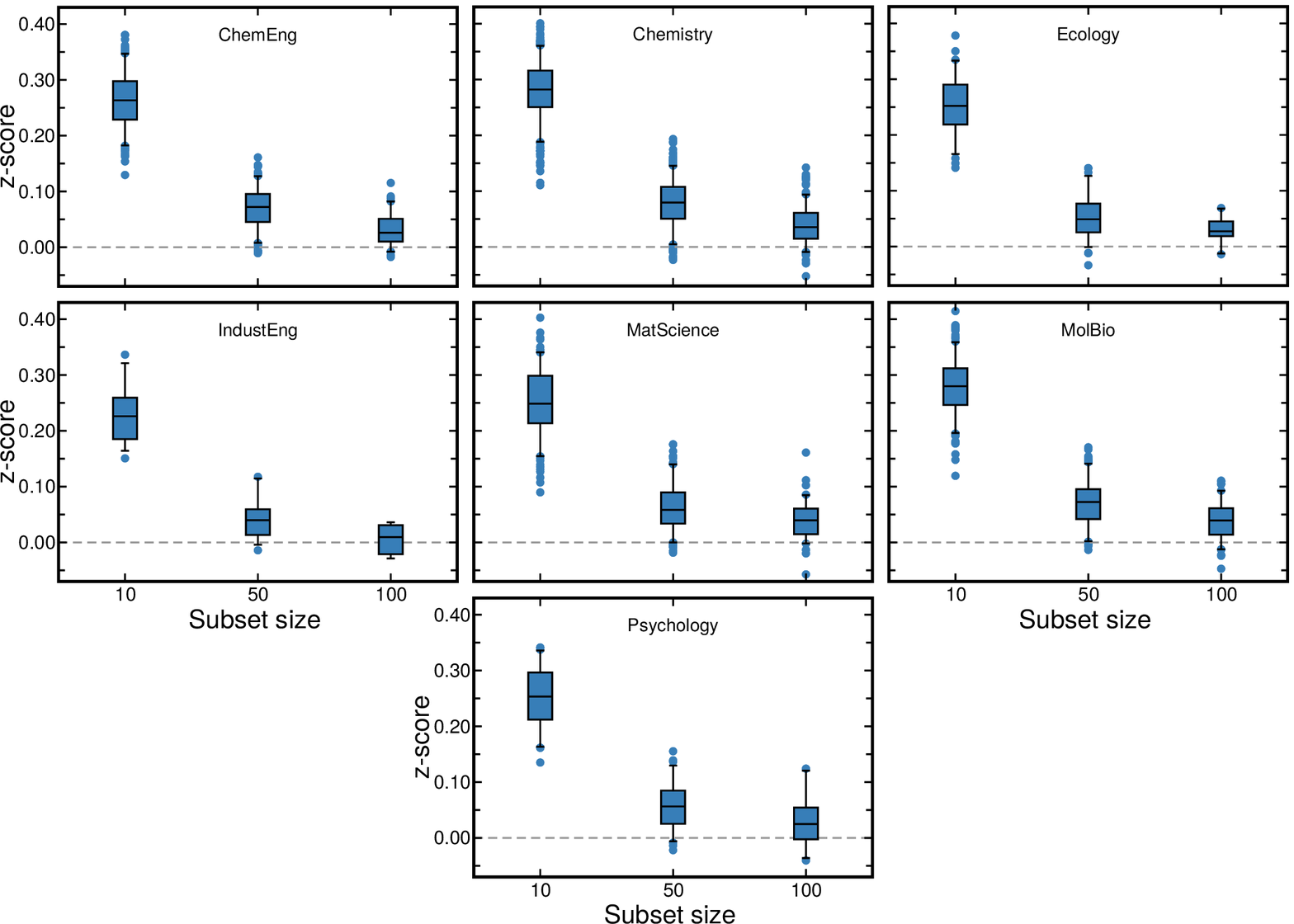}}
\caption{{\bf Dependence of $\hat{\sigma}$ estimates on number of publications at the individual level.} We use the same procedure as in Fig.~\ref{S1_Fig}, except here we show the results for the dependence of $\hat{\sigma}$ on sample size. Estimates of $\hat{\sigma}$ are more dependent of sample size than $\hat{\mu}$. However, as in the case of $\hat{\mu}$, the dependence of $\hat{\sigma}$ on sample size decays rapidly with increasing sample size. Researchers with $N_p <$ 100 are omitted from the calculation on the subset of size 100.}
\label{S2_Fig}
\end{figure}

Finally we estimate the parameters in Eg.~\eqref{eq:lognorm-eq} for chemistry journals and compare $\hat{\mu}$ of chemistry departments and journals in selected years, and all chemistry researchers in our database (Fig.~\ref{fig7}. See Fig.~\ref{S4_Fig} for $\hat{\sigma}$ and $f_s$ comparison). In order to make sense of this comparison, we must note a few aspects about the data. The researchers in the database were affiliated with the top 30 chemistry departments in the U.S., whereas the set of chemistry departments covers all the chemistry departments from very high research activity universities. Thus, it is natural that the typical $\hat{\mu}$ of researchers is higher than that of departments. Not surprisingly, we find that $\hat{\mu}$ is typically the lowest for journals.

\section*{Discussion}

The ever-growing size of the scientific literature precludes researchers from following all developments from even a single sub-field. Therefore researchers need proxies of quality in order to identify which publications to browse, read, and cite. Three main heuristics are familiar to most researchers: institutional reputation, journal reputation, and author reputation.

Author reputation has the greatest limitations. Researchers are not likely to be known outside their (sub-)field and young researchers will not even be known outside their labs. Similarly, if we exclude a few journals with multidisciplinary reputations (Nature, Science, PNAS, NEJM), the reputation of a scientific journal is unlikely to extend outside its field. Institutional reputations are the most likely to be known broadly. Cambridge, Harvard, Oxford, and Stanford are widely recognized. However, one could argue that institutional reputation is not a particularly useful heuristic for finding quality publications within a specific research field.

Our results show that the expected citability of scientific publications published by (i) the researchers in a department, (ii) a given scientific journal, or (iii) a single researcher can be set on the single scale defined by $\mu$. Thus, for a researcher whose publications are characterized by a very high $\mu$, authorship of a publication may give a stronger quality signal about the publication than the journal in which the study is being published. Conversely, for an unknown researcher the strongest quality signal is likely to be the journal where the research is being published or the institution the researcher is affiliated with. Our results thus provide strong evidence for the validity of the heuristics used by most researchers and clarify the conditions under which they are appropriate.

\section*{Materials and Methods}

\subsection*{Model Fitting and Hypothesis Testing}
We estimate the discrete lognormal model parameters of Eq.~\eqref{eq:lognorm-eq} for all 1,283 researchers in our database using a maximum likelihood estimator \cite{Stringer2010}. We then test the goodness of the fit, at an individual level using the $\chi^2$ statistical test. We bin the empirical data in such a way that there are at least 5 expected observation per bin. To assess significance we calculate the $\chi^2_o$ statistic for each researcher and then, for each of them, re-sample their citation records using bootstrap (1,000 samples) and calculate a new value of the statistics $\chi^2_i$ ($i = $ 1 $, \dotsc,$ 1,000). We then extract a p-value by comparing the observed statistic $\chi^2_o$ with the re-sampled $\chi^2$ distribution. Finally we use a multiple hypothesis correction \cite{Benjamini1995}, with a \emph{false discovery rate} of 0.05, when comparing the model fits with the null hypothesis.

\subsection*{Generation of Theoretical Performance Indicators}
For each discipline we take the average value of $\hat{\sigma}$ and 20 equally spaced values of $\mu$ between 0.5 and 2.0. We then generate 1,000 datasets of 50 and 200 publications by random sampling from Eq.~\eqref{eq:lognorm-eq}. We then fit the model individually to these 2,000 synthetic datasets and extracted the \emph{h}-index, average number of citations, total number of citations and median number of citations to publications with at least one citation. Finally, for each value of $\mu$, we calculate the average and the 95\% confidence interval of all the indicators.

\subsection*{Manipulation Procedure for \emph{h}-index}

We try to increase the \emph{h}-index of a researcher by self-citations alone, i.e., we assume the researcher does not receive citations from other sources during this procedure. The procedure works by adding only the minimum required citations to those publications that would cause the \emph{h}-index to increase. Consider researcher John Doe who has 3 publications with $\{n_a\}$ = (2,2,5). Doe's \emph{h} is 2. Assuming those publications don't get cited by other researchers during this time period, to increase \emph{h} by 1, Doe needs to publish only one additional publication with two self-citations; to increase \emph{h} by 2 he must instead produce five publications with a total of eight self-citations, four of which to one of the additional five publications. We execute this procedure for all researchers in the database until they reached a \emph{h}-index of 100.

\subsection*{Manipulation Procedure for $\mu$}

The manipulation of $\mu$ is based on Eq.~\eqref{eq:mu-med}. We try to change a researcher's $\mu$ by increasing the median number of citations to publications which have at least one citation already. We consider only self-citations originating from secondary publications, i.e., publications that will not get cited. For a given corpus of publications we first define a target increase in median, $x$ and then calculate the number of self-citations needed to increase the current median by $x$ citations and the corresponding number of secondary publications. We then take the initial corpus of publications and attempt to increase the median citation by $x$ + 1. We repeat this procedure until we reach an increase in median citation of 2000.

\bibliographystyle{unsrtnat}
\bibliography{arxiv_moreira}

\clearpage
\beginsupplement

\clearpage

\clearpage

\begin{figure}[t]
\centerline{\includegraphics[width=0.9\textwidth]{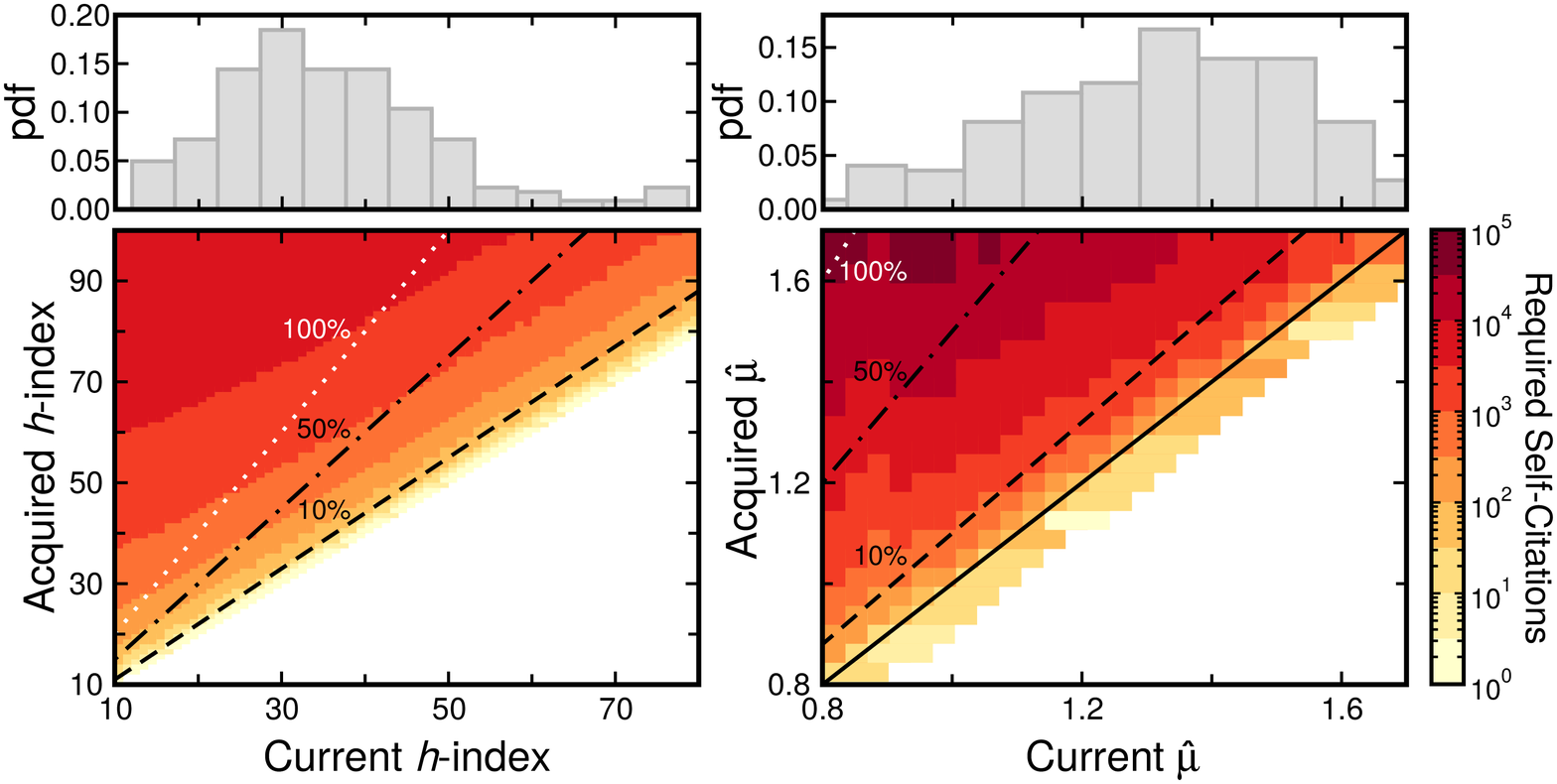}}
\caption{{\bf Susceptibility of impact measures to manipulation.} We used the same procedure as in Fig.~\ref{fig6}, except here we show the required number of publications with self-citations that researchers need to publish in order to increase their indicators. Other details are the same as in Fig.~\ref{fig6}.}
\label{S3_Fig}
\end{figure}

\clearpage

\begin{figure}[t]
\centerline{\includegraphics[scale=0.5]{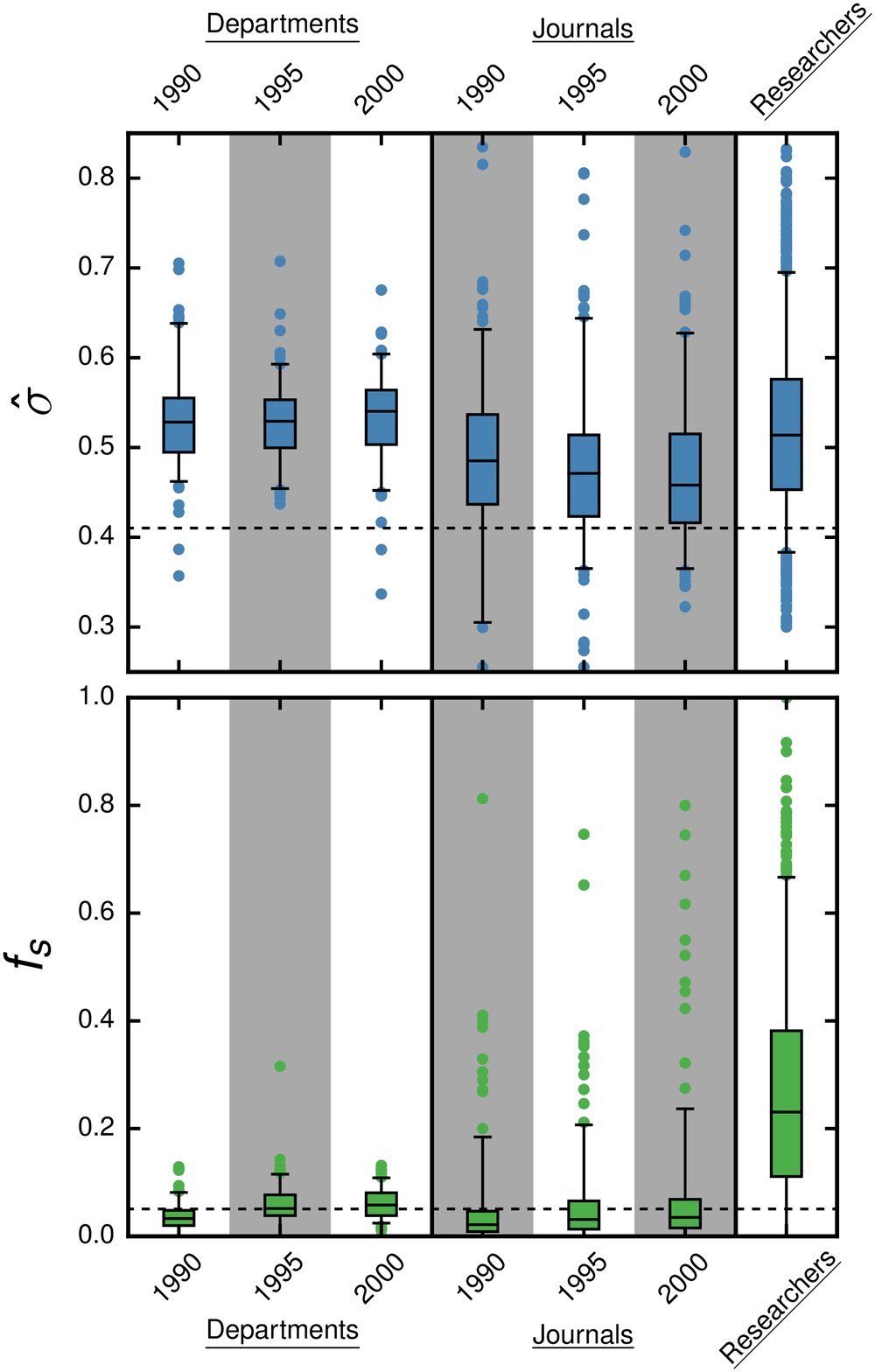}}
\caption{{\bf Comparison of $\hat{\sigma}$ and $f_s$ across departments, journals, and researchers.} We show the maximum likelihood fitted $\hat{\sigma}$ (\textbf{top}) and the fraction of secondary publications (\textbf{bottom}) for chemistry departments and chemistry journals in select years, and for all chemistry researchers in our database. The black horizontal dashed lines mark the value of the corresponding parameter for the \textit{Journal of the American Chemical Society} in 1995. For clarity, we do not show $\hat{\sigma}$ for 19 journals and 9 researchers that are outliers.}
\label{S4_Fig}
\end{figure}

\clearpage

\begin{table}[t]
\begin{center}
\begin{tabular}{clllll}
\hline
{\bf Parameter} & {\bf Discipline} & {\bf slope ($m$)} & {\bf intercept ($b$)} & {\bf R$^2$} & {\bf \textit{p}} \\
\hline
~ & ChemEng &  0.051 &  1.218 &  0.00187 &  0.5240 \\
~ & Chemistry &  0.073 &  1.286 &  0.00668 &  0.0744 \\
~ & Ecology &  -0.150 &  1.658 &  0.00844 &  0.4502 \\
$\hat{\mu} = m N_p + b$ & IndustEng &  0.379 &  0.348 &  0.04305 &  0.3166 \\
~ & MatScience &  0.106 &  1.043 &  0.01114 &  0.1278 \\
~ & MolBio &  0.156 &  1.332 &  0.01909 &  0.0410 \\
~ & Psychology &  0.104 &  1.229 &  0.00496 &  0.5732 \\
\hline
~& ChemEng &  0.067 &  0.379 &  0.03542 &  0.0052* \\
~ & Chemistry &  0.031 &  0.418 &  0.00650 &  0.0862 \\
~ & Ecology &  0.059 &  0.434 &  0.00806 &  0.4524 \\
$\hat{\sigma} = m N_p + b$  & IndustEng &  -0.094 &  0.764 &  0.01657 &  0.5380 \\
~ & MatScience &  0.033 &  0.502 &  0.00921 &  0.1598 \\
~ & MolBio &  0.064 &  0.415 &  0.01688 &  0.0592 \\
~ & Psychology &  -0.092 &  0.761 &  0.02026 &  0.2302 \\
\hline
\end{tabular}
\end{center}
\caption{{\bf Individual lognormal parameters show no dependence on $N_p$} For each researcher within each of the seven disciplines we perform least-squares linear regression between the lognormal parameters $\hat{\mu}$ and $\hat{\sigma}$, and $\textrm{log}_{10}(N_p)$. We used a permutation test to calculate the \textit{p}-values: for each set of pairs, $(\hat{\mu}, N_p)$ and $(\hat{\sigma}, N_p)$, we performed 10,000 random swaps of all $N_p$ and subsequent regression; we obtained a \textit{p}-value by comparing the original slope of the fit with the distribution of the permuted slopes. $^{*}p<$ 0.05/7 $\sim$ 0.0074.}
\label{S1_Table}
\end{table}

\clearpage

\begin{table}[t]
\begin{center}
\begin{tabular}{cllllll}
\hline
{\bf Parameter} & {\bf Discipline} & {\bf Mean} & {\bf Std Dev} & {\bf Min} & {\bf Median} & {\bf Max} \\
\hline
~ &  ChemEng &  1.354 &  0.256 &  0.566 &  1.320 &  1.921 \\
~ &  Chemistry &  1.439 &  0.225 &  0.689 &  1.437 &  2.179 \\
~ &  Ecology &  1.395 &  0.308 &  0.702 &  1.375 &  1.999 \\
$\hat{\mu}$ &  IndustEng &  1.012 &  0.313 &  0.603 &  1.046 &  1.466 \\
~ &  MatScience &  1.250 &  0.266 &  0.629 &  1.254 &  1.947 \\
~ &  MolBio &  1.624 &  0.253 &  0.950 &  1.641 &  2.250 \\
~ &  Psychology &  1.437 &  0.318 &  0.545 &  1.427 &  1.967 \\
\hline
~ &  ChemEng &  0.508 &  0.080 &  0.323 &  0.513 &  0.764 \\
~ &  Chemistry &  0.468 &  0.095 &   0.300 &  0.482 &  0.956 \\
~ &  Ecology &  0.536 &  0.124 &  0.359 &  0.546 &  0.896 \\
$\hat{\sigma}$ &  IndustEng &  0.604 &  0.150 &  0.438 &  0.591 &  0.796 \\
~ &  MatScience &  0.559 &  0.095 &  0.335 &  0.568 &  0.969 \\
~ &  MolBio &  0.525 &  0.105 &  0.362 &  0.541 &  1.006 \\
~ &  Psychology &  0.550 &  0.137 &  0.398 &  0.585 &  0.954 \\
\hline
\end{tabular}
\end{center}
\caption{\bf Individual discipline statistics of the lognormal model parameters}
\label{S2_Table}
\end{table}

\end{document}